# Mars Entry Trajectory Planning with Range Discretization and Successive Convexification

Xu Liu[1] and Shuang Li[2]

*Nanjing University of Aeronautics and Astronautics, Nanjing, 211106, China*

Ming Xin[3]

*University of Missouri, Columbia, MO 65211, USA*

## I. Introduction

Mars entry refers to the flight phase from entering the atmosphere to the deployment of deceleration devices, where the constraints of heat rate, pressure, and normal load should be met [1]. The reference trajectory tracking guidance has been widely adopted since constraints can be addressed in advance by trajectory planning [2, 3]. To ensure the safety of a Mars mission, it is essential to plan the entry trajectory that satisfies all constraints with high accuracy.

Recently, convex optimization has attracted increasing attention of researchers to solve aerospace trajectory optimization problems due to its characteristics of polynomial time convergence and theoretical global optimality [4]. The pioneering lossless convexification technique was proposed by Acikmese's team [5], and this approach was improved to take into account more complex physical factors, including the minimum-landing-error objective, planetary self-rotation effect and state-triggered constraint in more realistic problems [6-9]. For the atmospheric entry problem, Liu et al. utilized sine and cosine of the bank angle as new control inputs and introduced additional second-order cone inequality constraint to replace the nonconvex equality constraint on the new controls [10, 11], which can convert the constrained entry problem into a convex one and effectively suppress the high-frequency jitters of the bank angle. Then the convex problem was solved by linearization, discretization, and successive convexification techniques with the well-established interior point method. Owing to the merits of fast convergence and low computation complexity, this strategy developed by Liu et al. nearly became the standard approach to addressing entry trajectory optimization problem via sequential convex programming (SCP). Inspired by this scheme, Zhao and Song [12] proposed a multi-phase convex programming algorithm for entry trajectory planning by means of normalized lift coefficient and waypoints optimization. Wang and Grant [13] improved the work in [10, 11] by using the bank angle rate as the new control input, which decouples the state and control in entry dynamics so as to eliminate the high-frequency jitters in a simpler way. Furthermore, the backtracking line-search and dynamic trust-region techniques were introduced by Wang [14, 15]. Sagliano and Mooij [16] proposed the pseudospectral convex programming approach, where the optimal control problem is discretized at flip-Radau collocation points and then solved in the framework of second-order cone programming (SOCP). Lately, Zhou et al. [17] investigated the atmospheric entry trajectory planning problem with SCP and an adaptive mesh refinement technique.

---
[1] Ph.D. Candidate, Department of Aerospace Control Engineering, Email: liuxujx@nuaa.edu.cn.
[2] Corresponding Author, Professor, Department of Aerospace Control Engineering, Email: lishuang@nuaa.edu.cn.
[3] Professor, Department of Mechanical and Aerospace Engineering, Email: xin@missouri.edu, Associate Fellow, AIAA.

Most existing literatures study the optimal control problem with a fixed final time rather than the free final time. To address this issue, Refs. [5, 18, 19] employed a bi-level optimization strategy to find the optimal final time. Another commonly used method is to transform the time interval $[0, t_f]$ into a normalized interval $[-1, 1]$ in the pseudospectral transcription paradigm [16, 20, 21] or into the interval $[0, 1]$ in the classical optimal control framework [9, 12]. A new method to solve the optimal control problem (OCP) with free final time is to choose a new independent variable for dynamics formulation. In Refs. [22, 23], the altitude was used as an independent variable for trajectory optimization of aerodynamically controlled missiles and powered descent and landing. In [24], the transfer angle was adopted to rewrite the Earth-to-Mars low-thrust planar transfer dynamics.

In Ref. [10], the first-order linearization approximation technique was used to convexify the integral performance index, which in turn introduces linearization error into the performance index. Moreover, artificial infeasibility may occur during the successive optimization due to an improper linearization. To mitigate this issue, Mao [25] and Szmuk [9] introduced the virtual control and trust-region techniques to compensate the linearized dynamics. Wang [15, 26] further improved the trust-region algorithm with adaptive techniques, where the trust-region radius was refined by evaluating the exact decrease of the performance index.

In this note, a range discretization-based sequential convex programming (r-SCP) approach is proposed. The main contributions can be summarized as follows: 1) The Mars entry trajectory planning problem is reformulated by using the range of entry trajectory as an independent variable rather than the energy or time. This can avoid numerical inaccuracy of dynamics propagation that occurs if the problem is evenly discretized with respect to the independent variable energy, especially in the high altitude. 2) The minimum-time performance index is convexified by taking the flight time as one of the state variables; and the constraints on bank angle along with its rate are addressed by introducing a new control input and inequality relaxation. These two methods suggest an effective strategy to deal with non-convex state variables or control input constraints, that is, to convexify constraints by change of variables, inequality relaxation, or state-space augmentation. 3) With these new formulations, range discretization and successive convexification approaches are applied for Mars entry trajectory planning in the existing convex optimization framework with virtual control and adaptive trust-region techniques. Furthermore, performance comparison of the proposed algorithm with some classical optimization methods is conducted for the entry guidance problem.

## II. Mars Entry Trajectory Planning Problem Formulation

### A. Choice of new independent variable

It is common to use the non-dimensional flight time $\tau = t/t_s$ or negative specific mechanical energy (abbreviated as energy in the following) $e = 1/r - V^2/2$ as an independent variable in entry trajectory optimization problems [10, 13], where $t$ is the flight time and $t_s = \sqrt{R_0/g_0}$; $R_0$ and $g_0$ denote Mars reference radius and gravitational acceleration at $R_0$, respectively. $r$ is the dimensionless radial distance from Martian center to the vehicle normalized by $R_0$. The relative velocity $V$ is normalized by velocity scaling coefficient $V_s = \sqrt{R_0 g_0}$. Since the flight time is not a typical concern during entry mission as it is not related to boundary and path constraints, the energy is preferred to time as an independent variable. The energy monotonically increases when the effect of self-rotation is ignored:

$$\frac{de}{d\tau} = DV > 0 \tag{1}$$

where $D$ is the non-dimensional drag and will be defined later. Nevertheless, the derivative $de/d\tau$ will no longer be in this simple form if the self-rotation-dependent terms are considered:

$$\frac{de}{d\tau} = DV - \Omega^2 rV \cos\phi \left(\cos\phi \sin\gamma - \sin\phi \cos\gamma \cos\psi\right) \tag{2}$$

where $\Omega$ is the Mars self-rotation rate non-dimensionalized by $t_s$. $\phi$ and $\gamma$ are the latitude and fight-path angle, respectively. $\psi$ is the heading angle measured clockwise in the local horizontal plane from the north. As can be seen from Eq. (2), the modeling accuracy of dynamics with respect to the energy variable is not guaranteed if the self-rotation-dependent term is ignored. To demonstrate this, propagation results of the two dynamics with respect to energy and time are listed in Table 1, in which $h$ is the flight altitude and $s$ denotes the range traveled by the vehicle after entry defined by $ds/d\tau = V \cos\gamma$.

**Table 1 Numerical integration results of entry dynamics with time and energy as independent variables**

| | Initial value | Final value | |
|---|---|---|---|
| | | Time | Energy |
| $h$ (m) | 125000 | 16232.04 | 16203.80 |
| $\theta$ (deg) | -90 | -71.9758 | -71.9794 |
| $\phi$ (deg) | -45 | -41 | -41.1441 |
| $V$ (m/s) | 5500 | 752.6682 | 724.6368 |
| $\gamma$ (deg) | -13.5 | -11.4325 | -11.4560 |
| $\psi$ (deg) | 85 | 48.2555 | 48.2447 |
| $\sigma$ (deg) | -50 | -50 | -50 |
| $s$ (m) | 0 | 830182.57 | 829997.33 |

As shown in Table 1, the final altitude integrated by using energy-based dynamics is about 28.24 m lower than the one using the time-based dynamics. The terminal difference in terms of parachute deployment longitude and latitude is about $[-3.6281, -1.444] \times 10^{-3}$ deg or $[-215.12, -85.63]$ m, which exceeds the tolerable 100 m error for future human-scale Mars mission. Consequently, the energy-based model for Mars entry is not accurate enough. Moreover, the flight altitude $h$ decreases dramatically at the beginning with a large magnitude of flight path angle because the thin atmosphere in high altitude cannot provide sufficient deceleration. By contrast, the energy changes slowly. Thus, the altitude-energy profile is featured by a steep slope during the initial phase of entry. In order to ensure numerical integration accuracy, the step size should be set small enough to capture such a rapid dynamic behavior if the energy is evenly discretized. Accordingly, the number of discretization nodes and the scale of optimization increase significantly. However, a large-scale optimization problem is tricky. Moreover, modeling with respect to energy also requires a fixed final altitude and speed to determine the final energy. However, the altitude and speed are usually constrained within a parachute box rather than constants in Mars entry.

It is worth noting that the range $s$ increases monotonically since its derivative $ds/d\tau = V \cos\gamma$ is positive during the whole entry mission. In addition, the derivative of the range is simple enough to be an alternative for replacing the approximate derivative of energy. Thus, the Mars entry dynamics can be reformulated without losing accuracy by using the range as an independent variable. Moreover, the range to go $s_{togo} = \int_0^{\tau_f} -V\cos\gamma/r d\tau$ (or $\bar{s} = \int_0^{\tau_f} V\cos\gamma/r d\tau$) [27] can also be chosen as the independent variable because it has the same monotonicity property and its initial and final boundaries are specified in an entry mission with the given target latitude and longitude.

## B. Dynamics and Constraints

The equations of motion can be rewritten with respect to $s$ as follows:

$$\frac{dr}{ds} = \tan\gamma \tag{3}$$

$$\frac{d\theta}{ds} = \frac{\sin\psi}{r\cos\phi} \tag{4}$$

$$\frac{d\phi}{ds} = \frac{\cos\psi}{r} \tag{5}$$

$$\frac{dV}{ds} = -\frac{D}{V\cos\gamma} - \frac{\tan\gamma}{r^2 V} + \frac{\Omega^2 r \cos\phi(\cos\phi\tan\gamma - \sin\phi\cos\psi)}{V} \tag{6}$$

$$\frac{d\gamma}{ds} = \frac{L\cos\sigma}{V^2\cos\gamma} - \frac{1}{r^2 V^2} + \frac{1}{r} + \frac{2\Omega\cos\phi\sin\psi}{V\cos\gamma} + \frac{r\Omega^2\cos\phi(\cos\phi + \cos\psi\sin\phi\tan\gamma)}{V^2} \tag{7}$$

$$\frac{d\psi}{ds} = \frac{L\sin\sigma}{V^2\cos^2\gamma} + \frac{\sin\psi\tan\phi}{r} + \frac{2\Omega(\sin\phi - \cos\phi\tan\gamma\cos\psi)}{V\cos\gamma} + \frac{\Omega^2 r \sin\phi\cos\phi\sin\psi}{V^2\cos^2\gamma} \tag{8}$$

where $\theta$ is the longitude and $\sigma$ is the bank angle measured positive to the right from the view inside the vehicle. The non-dimensional drag and lift accelerations $D$ and $L$ are given by:

$$D = R_0\rho V^2 S_r C_D/(2m), \quad L = R_0\rho V^2 S_r C_L/(2m) \tag{9}$$

where $\rho$ is the atmospheric density, $S_r$ denotes the reference area, $m$ is the mass of the unpowered vehicle. $C_L$ and $C_D$ are drag and lift coefficients of the vehicle, respectively. The Mars atmospheric density model is approximated by an exponential function of altitude as:

$$\rho = \rho_0 e^{-R_0(r-1)/h_s} \tag{10}$$

where $\rho_0$ is atmospheric density on the planetary surface, and $h_s$ is the scale altitude. The Mars entry vehicle keeps a nominal trim angle of attack and a desired $C_L/C_D$ via offsetting its center of gravity. Therefore, the $C_L$ and $C_D$ nearly remain unchanged in the hypersonic flight, and the bank angle $\sigma$ is the only control variable.

Three typical path constraints during Mars entry are heat rate $\dot{Q}$ with respect to the dimensional time $t$, dynamic pressure $q$, and normal load $a$:

$$\dot{Q} = k_Q\sqrt{\rho/R_n}(VV_s)^{3.15}, \quad q = \rho(VV_s)^2/2 \leq q_{max}, \quad a = \sqrt{L^2 + D^2} \leq a_{max} \tag{11}$$

where $k_Q$ is the heat rate related coefficient, $R_n$ is vehicle's nose radius, $\dot{Q}_{max}$, $q_{max}$ and $a_{max}$ are their tolerable maximum values.

The initial conditions and desired final conditions are assumed to be specified:

$$[r(0),\theta(0),\phi(0),V(0),\gamma(0),\psi(0)]^T = [r_0,\theta_0,\phi_0,V_0,\gamma_0,\psi_0]^T \tag{12}$$

$$\boldsymbol{E}[r(s_f),\theta(s_f),\phi(s_f),V(s_f),\gamma(s_f),\psi(s_f)]^T = \boldsymbol{E}[r_f,\theta_f,\phi_f,V_f,\gamma_f,\psi_f]^T \tag{13}$$

where $(\circ)_0$ and $(\circ)_f$ denote the initial and final values of non-dimensional variables, respectively, and $\tau_f$ is the non-dimensional final time, $\boldsymbol{E}$ is a diagonal matrix with diagonal entries being 0 and 1 to determine the final boundary constraints according to different performance indices.

Constraints on the bank angle are given by:

$$|\sigma| \leq \sigma_{max}, \quad |\dot{\sigma}| \leq \dot{\sigma}_{max} \tag{14}$$

where $\dot{\sigma}$ is the bank angle rate with respect to the dimensional time $t$. The operator $(\circ)_{min}$ and $(\circ)_{max}$ denote the corresponding lower and upper bound of a variable, respectively.

## C. Reformulated optimal control problem

For the atmospheric entry problem, the following performance indices are generally considered:

$$J_1 = -r(s_f), \quad J_2 = V(s_f), \quad J_3 = \int_0^{s_f} \frac{1}{V\cos\gamma} ds \tag{15}$$

which represent maximizing final altitude, minimizing the speed or flight time at the instant of parachute deployment, respectively. Note that the final range $s_f$ is an unconstrained variable, which leads to the difficulty in solving an optimal control problem in the framework of mathematical programming. To circumvent this issue, the final range is regarded as a dilation coefficient to normalize the interval from $[0, s_f]$ into $[0,1]$ before linearization, which is adopted in [9]. Then, this problem can be viewed as a fixed-final-range optimization problem because the final normalized dimensionless range is always equal to one. Following the process outlined in [9], one can obtain the derivative according to the chain rule:

$$x' \triangleq \frac{d\boldsymbol{x}(S)}{dS} = \frac{d\boldsymbol{x}(s)}{ds}\frac{ds}{dS} = \frac{d\boldsymbol{x}(s)}{ds}s_f = s_f \boldsymbol{F}(\boldsymbol{x}(S)), \quad s/s_f \triangleq S \in [0,1] \tag{16}$$

where $\boldsymbol{x} = [r, \theta, \phi, V, \gamma, \psi]^T \in \mathbb{R}^6$, and $\boldsymbol{F}(\boldsymbol{x}(S))$ consists of the right-hand side terms in Eqs. (3)~(8). Then, the final boundary constraints in Eq. (13) and performance indices in Eq. (15) are rewritten as:

$$\boldsymbol{E}_i[r(1), \theta(1), \phi(1), V(1), \gamma(1), \psi(1)]^T = \boldsymbol{E}_i[r_f, \theta_f, \phi_f, V_f, \gamma_f, \psi_f]^T, \quad i = 1,2,3 \tag{17}$$

$$J_1 = -r(1), \quad J_2 = V(1), \quad J_3 = \int_0^1 s_f \frac{1}{V\cos\gamma} dS \tag{18}$$

where $\boldsymbol{E}_1 = diag([0,1,1,0,0,0])$, $\boldsymbol{E}_2 = \boldsymbol{E}_3 = diag([1,1,1,0,0,0])$ make the final boundary constraints match each of the three performance indices, respectively.

Thus far, the constrained nonlinear optimal trajectory planning problem for Mars entry can be formulated as:

$$\text{P0}: \min_{\boldsymbol{x}, u, s_f} \text{Eq. (18)}$$

$$\text{subject to} \quad \text{Eqs. (11), (12), (14), (16), (17)}$$

### III. Convexification Formulation

In this section, the non-convex fixed-final-range trajectory planning problem is converted into a SOCP problem by linearization, then the problem is solved via successive convexification.

#### A. Linearization of dynamics and constraints, and performance convexification

As reported in Refs. [10, 13], high-frequency jitters will occur due to the coupling between states and control input after linearization of the nonlinear entry dynamics with respect to $\sigma$. Therefore, it is necessary to modify the dynamics before linearization. To address the issue of coupling, the method in Ref. [13] is employed by introducing a new control variable $u$ and augmenting the bank angle $\sigma$ into the state vector. The redefined control input $u$ with respect to the new state variable $\sigma$ is given by:

$$u = s_f^{-1}\frac{d\sigma}{dS} \tag{19}$$

Accordingly, the constraint defined in Eq. (14) can be relaxed into the following form:

$$|\sigma| \leq \sigma_{\max}, \quad |\dot{\sigma}| = \left|\frac{d\sigma}{dt}\right| = \left|\frac{d\sigma}{dS}\frac{dS}{ds}\frac{ds}{d\tau}\frac{d\tau}{dt}\right| = \left|us_f \frac{1}{s_f} V \cos\gamma \frac{1}{t_s}\right| = \left|\frac{uV\cos\gamma}{t_s}\right| \leq |u|\frac{V_{\max}}{t_s} \leq \dot{\sigma}_{\max} \qquad (20)$$

Note that a relaxation is used to transform this nonlinear inequality into a linear one since $V\cos\gamma \leq V_{\max}$ holds, and the predetermined constant $V_{\max} = 6000/V_s$ is the maximum speed that a vehicle can fly in a Mars entry mission. This relaxation guarantees the maximum control input meets the constraint on the bank angle rate.

Furthermore, a convex performance index is necessary for the successive convexification detailed in the next subsection. Note that the time-optimal performance index shown in Eq. (18) is a non-convex one. Generally, the first-order Taylor series expansion is used to deal with the performance index with a nonlinear integrant. However, this relaxation method introduces the approximation error. In this paper, a new and simpler method is proposed. To this end, the dimensionless time $\tau$ is incorporated into the state vector, and its derivative is specified as:

$$\frac{d\tau}{dS} = s_f \frac{1}{V\cos\gamma} \qquad (21)$$

Then, the state vector augmented by $\tau$ becomes:

$$\boldsymbol{x} = [r, \theta, \phi, V, \gamma, \psi, \sigma, \tau]^T \in \mathbb{R}^8 \qquad (22)$$

and the new equivalent minimum-time performance index becomes $J_3 = \tau(1)$. As such, the decoupled dynamics with a convex performance index is obtained, and the reformulated dynamics can be rewritten as follows:

$$\boldsymbol{x}' = s_f(\boldsymbol{f}(\boldsymbol{x}) + \boldsymbol{B}u + \boldsymbol{f}_\Omega(\boldsymbol{x})) \qquad (23)$$

where $\boldsymbol{B} = [0,0,0,0,0,0,1,0]^T$; $\boldsymbol{f}(\boldsymbol{x}) \in \mathbb{R}^8$ and $\boldsymbol{f}_\Omega(\boldsymbol{x}) \in \mathbb{R}^8$, respectively, denote the terms without and with the self-rotation on the right-hand side of Eqs. (3)~(8), (19) and (21).

In order to solve the entry trajectory planning problem in a convex optimization framework, the non-convex dynamics and constraints can be relaxed into linear ones by the first-order Taylor series approximation at the point $(x^k, u^k, s_f^k)$ of a reference trajectory, which is obtained at the $k$-th iteration in the successive convexification procedure discussed in Section III.C.

$$x' = s_f \left[ f(x) + Bu + f_\Omega(x) \right] \approx s_f^k \left. \frac{\partial f(x)}{\partial x} \right|_{x^k, u^k} (x - x^k) + s_f^k \left. \frac{\partial f_\Omega(x)}{\partial x} \right|_{x^k, u^k} (x - x^k) + s_f^k B(u - u^k)$$
$$+ \left[ f(x^k) + Bu^k + f_\Omega(x^k) \right] (s_f - s_f^k) + s_f^k \left[ f(x^k) + Bu^k + f_\Omega(x^k) \right] = A^k x + B^k u + c^k s_f + d^k \quad (24)$$

where $A^k = s_f^k \left. \frac{\partial f}{\partial x} \right|_{x^k, u^k}$, $B^k = s_f^k B$, $c^k = f(x^k) + Bu^k + f_\Omega(x^k)$, and $d^k = -s_f^k \left. \frac{\partial f}{\partial x} \right|_{x^k, u^k} x^k - s_f^k B u^k$. It should be noted that the first-order term $s_f^k \left. \frac{\partial f_\Omega(x)}{\partial x} \right|_{x^k, u^k} (x - x^k)$ in Eq. (24) is omitted since its Euclidean norm is much smaller than the other term $s_f^k f_\Omega(x^k)$.

After the above steps, the path constraint is the only source of non-convexity. Analogous to the linearization of dynamics, the path constraint can be relaxed as the following form:

$$f_i \approx f_i(r^k, V^k) + f_i'(r^k, V^k) \cdot [r - r^k, V - V^k]^T \leq f_{i,\max} \quad \text{for } i = 1, 2, 3 \quad (25)$$

where $f_1 \triangleq \dot{Q}$, $f_2 \triangleq q$, $f_3 \triangleq n$, $f_{1,\max} \triangleq \dot{Q}_{\max}$, $f_{2,\max} \triangleq q_{\max}$, $f_{3,\max} \triangleq a_{\max}$ and

$$f_1'(r^k, V^k) = \left[ \frac{\partial f_1}{\partial r}, \frac{\partial f_1}{\partial V} \right]_{r^k, V^k} = \left[ -0.5 R_0 k_Q (V V_s)^{3.15} \sqrt{\rho} / h_s, 3.15 k_Q V_s^{3.15} V^{2.15} \sqrt{\rho} \right]_{r^k, V^k} \quad (26)$$

$$f_2'(r^k, V^k) = \left[ \frac{\partial f_2}{\partial r}, \frac{\partial f_2}{\partial V} \right]_{r^k, V^k} = \left[ -0.5 R_0 \rho (V V_s)^2 / h_s, V_s \rho V \right]_{r^k, V^k} \quad (27)$$

$$f_3'(r^k, V^k) = \left[ \frac{\partial f_3}{\partial r}, \frac{\partial f_3}{\partial V} \right]_{r^k, V^k} = \left[ -0.5 R_0^2 S_r (\sqrt{C_L^2 + C_D^2}) \rho V^2 / m h_s, R_0 S_r (\sqrt{C_L^2 + C_D^2}) \rho V / m \right]_{r^k, V^k} \quad (28)$$

Now all the equality and inequality constraints as well as the performance index are convex, then the entry trajectory planning problem can be solved in the framework of convex programming.

To avoid the artificial infeasibility and unbounded relaxation, an additional control input $v$ called virtual control [25, 28] and the adaptive trust-region technique [9] are employed. Specifically, the virtual control proposed in [25, 28] is introduced into linearized dynamics to compensate the linearization error, thereby preventing the artificial infeasibility. Additionally, the trust-region constraint is enforced on state variables to avoid unbounded relaxation, where the adaptive trust-region radius introduced in [9] is used to adaptively determine the step size in each iteration and reduce the total number of optimization steps. The modified dynamics and augmented constraints are given by:

$$\boldsymbol{x}' = s_f(\boldsymbol{f}(\boldsymbol{x}) + \boldsymbol{B}\boldsymbol{u} + \boldsymbol{f}_\Omega(\boldsymbol{x})) \approx \boldsymbol{A}^k \boldsymbol{x} + \boldsymbol{B}^k \boldsymbol{u} + \boldsymbol{c}^k s_f + \boldsymbol{d}^k + \boldsymbol{v} \tag{29}$$

$$\left\| \boldsymbol{x} - \boldsymbol{x}^k \right\|_2 \leq \delta_x, \quad \left| s_f - s_f^k \right| \leq \delta_s \tag{30}$$

where the virtual control input $\boldsymbol{v} \in \mathbb{R}^8$ together with relaxation boundaries $\delta_x, \delta_s \in \mathbb{R}_+$ are new variables to be optimized, and the inequality involving the $l_2$-norm in Eq. (30) is a convex cone constraint that can be solved via SOCP. In turn, the performance index is accordingly modified by adding the penalty terms corresponding to the virtual control, trust-region and final range:

$$J = w_i J_i + w_v \left\| \boldsymbol{v} \right\|_1 + w_x \delta_x + w_s \delta_s, \quad i = 1, 2, 3 \tag{31}$$

where $w_i, w_v, w_x, w_s \in \mathbb{R}_+$ are the weight parameters prescribed by users. Similar augmented performance indices can also be found in [29, 30]. Among the three additional performance indices, the $l_1$-norm is adopted on $\boldsymbol{v}$ since it enables the virtual control to work only on discretization nodes with large errors, and not to compensate on others. As a result, less virtual control inputs are necessary when the algorithm converges. Note that $l_2$-norm can be used as well [30], but it makes these virtual controls converge to zero uniformly, suggesting that more variables and calculations are involved in $l_2$-norm optimization. The other terms about the trust-region and final range guarantee the bounded relaxation on state variables and the convergence of final range, respectively. The robustness of SCP is greatly improved by the virtual control and adaptive trust-region techniques. It is worth noting that these advantages are achieved at the expense of some optimality due to the augmented performance index. More discussions on this feature will be detailed based on the simulation results in Section IV.

## B. Range discretization

After linearization, it is still a continuous-time optimal control problem. To obtain the numerical solution, this problem is discretized at $N+1$ evenly distributed points over the interval $[0,1]$, and constraints are imposed at each point. The discretization points set is defined as $\{S_0, S_1, \cdots, S_N\}$ with a step $\Delta S = 1/N$ and $S_{n+1} = S_n + \Delta S$, $n = 0, 1, \cdots, N-1$. Correspondingly, the discretized state and control vectors at the $k$-th iteration are denoted as $x_n^k \triangleq x^k(S_n)$, and $u_n^k \triangleq u^k(S_n)$, Then. the dynamics can be propagated numerically via the trapezoidal rule over the reference trajectory $\{x_0^k, x_1^k, \cdots, x_N^k, u_0^k, u_1^k, \cdots, u_N^k, s_f^k\}$:

$$x_{n+1} = x_n + \frac{\Delta S}{2}(A_n^k x_n + B_n^k u_n + c_n^k s_f + d_n^k + v_n + A_{n+1}^k x_{n+1} + B_{n+1}^k u_{n+1} + c_{n+1}^k s_f + d_{n+1}^k + v_{n+1}) \quad (32)$$

where $A_n^k = s_f^k \left.\frac{\partial f}{\partial x}\right|_{x_n^k, u_n^k}$, $B_n^k = B_{n+1}^k = s_f^k[0,0,0,0,0,0,1,0]^T$, $c_n^k = f(x_n^k) + Bu_n^k + f_\Omega(x_n^k)$, and $d_n^k = -s_f^k \left.\frac{\partial f}{\partial x}\right|_{x_n^k, u_n^k} x_n^k - s_f^k B u_n^k$. Then Eq. (32) can be further rewritten as:

$$H_n x_n + H_{n+1} x_{n+1} + \frac{\Delta S}{2} B^k (u_n + u_{n+1}) + \frac{\Delta S}{2}(v_n + v_{n+1}) + G_{n+1} s_f = -\frac{\Delta S}{2}(d_n^k + d_{n+1}^k) \quad (33)$$

where $H_n = \left(I + \frac{\Delta S}{2} A_n^k\right)$, $H_{n+1} = -\left(I - \frac{\Delta S}{2} A_{n+1}^k\right)$, $G_{n+1} = \frac{\Delta S}{2}(c_n^k + c_{n+1}^k)$.

For notational convenience, we concatenate all decision variables $x_n$, $u_n$, $s_f$, and $v_n$ into a lumped vector $y = [x_0; x_1; \cdots; x_N; u_0; u_1; \cdots; u_N; v_0; v_1; \cdots; v_N; s_f] \in \mathbb{R}^{17(N+1)+1}$, and the dynamics is rewritten as a linear equality constraint on $y$:

$$[M_1 \quad M_2 \quad M_3 \quad M_4] y = P \quad (34)$$

$$M_1 = \begin{bmatrix} I & 0 & 0 & \cdots & 0 & 0 \\ H_0 & H_1 & 0 & \cdots & 0 & 0 \\ \vdots & \vdots & \vdots & \ddots & \vdots & \vdots \\ 0 & 0 & 0 & \cdots & H_{N-1} & H_N \end{bmatrix}, \quad M_2 = B^k M_3, \quad M_3 = \frac{\Delta S}{2}\begin{bmatrix} 0 & 0 & 0 & \cdots & 0 & 0 \\ I & I & 0 & \cdots & 0 & 0 \\ \vdots & \vdots & \vdots & \ddots & \vdots & \vdots \\ 0 & 0 & 0 & \cdots & I & I \end{bmatrix}, \quad M_4 = \begin{bmatrix} 0 \\ G_1 \\ \vdots \\ G_N \end{bmatrix} \quad (35)$$

$$P = -\frac{\Delta S}{2}\begin{bmatrix} -\frac{2}{\Delta S} x_0 \\ d_0^k + d_1^k \\ \vdots \\ d_{N-1}^k + d_N^k \end{bmatrix} \quad (36)$$

All constraints and modified performance index can be reformulated as:

$$f_{i,j} \leq f_{i,\max}, \quad |\sigma_j| \leq \sigma_{\max}, \quad |u_j| \leq \dot{\sigma}_{\max} t_s / V_{\max} \quad (37)$$

$$x(0) = x_0, \quad E_i x_N = E_i x(1) \quad (38)$$

$$\left\| \boldsymbol{x}_j - \boldsymbol{x}_j^k \right\|_2 \leq \delta_{x,j}, \quad \left| s_f - s_f^k \right| \leq \delta_s \tag{39}$$

$$J = w_i J_i + w_v \left\| \bar{\boldsymbol{v}} \right\|_1 + w_x \left\| \bar{\boldsymbol{\delta}}_x \right\|_2 + w_s \delta_s \tag{40}$$

where $i = 1, 2, 3$, $j = 0, 1, \cdots, N$, $\bar{\boldsymbol{\delta}}_x = [\delta_{x,0}; \delta_{x,1}; \cdots; \delta_{x,N}]$, $\bar{\boldsymbol{v}} = [\boldsymbol{v}_0; \boldsymbol{v}_1; \cdots; \boldsymbol{v}_N]$ and the trust-region constraint is a second-order cone. Note that decision variables are $\boldsymbol{y}$, $\bar{\boldsymbol{\delta}}_x$ and $\delta_s$ for the convexified optimal control problem.

### C. Successive convexification

With convexified dynamics, constraints and performance index, the r-SCP algorithm is summarized below:

1) Let $k = 0$ and initialize the state vector with $[r_0, \theta_0, \phi_0, V_0, \gamma_0, \psi_0, \sigma_0, \tau_0]$. Propagate the entry dynamics with a constant control input to obtain the initial guess trajectory $\{\boldsymbol{x}_0^0, \boldsymbol{x}_1^0, \cdots, \boldsymbol{x}_N^0, u_0^0, u_1^0, \cdots, u_N^0, s_f^0\}$.

2) For $k \geq 1$, solve the following optimal control problem to find the optimal decision variable $\boldsymbol{y}^k$ (or optimal trajectory $\{\boldsymbol{x}_0^k, \boldsymbol{x}_1^k, \cdots, \boldsymbol{x}_N^k, u_0^k, u_1^k, \cdots, u_N^k, s_f^k\}$ and virtual control $[\boldsymbol{v}_0^k; \boldsymbol{v}_1^k; \cdots; \boldsymbol{v}_N^k]$) with the previous solution $\{\boldsymbol{x}_0^{k-1}, \boldsymbol{x}_1^{k-1}, \cdots, \boldsymbol{x}_N^{k-1}, u_0^{k-1}, u_1^{k-1}, \cdots, u_N^{k-1}, s_f^{k-1}\}$ as the guess.

$$\text{P1:} \quad \min_{\boldsymbol{y}, \bar{\boldsymbol{\delta}}_x, \delta_s} \text{Eq. (40)}$$

subject to Eqs. (34), (37)~(39)

3) If $\left\| \bar{\boldsymbol{\delta}}_x \right\|_2 \leq \delta_{tot}$ and $\left\| \bar{\boldsymbol{v}} \right\|_1 \leq v_{tot}$, go to Step 4), otherwise set $k = k + 1$ and go to Step 2). Note that $\delta_{tot}$ and $v_{tot} \in \mathbb{R}_+$ are the given convergence tolerance specified by users.

4) The solution of the Mars entry trajectory planning problem is found to be $\{\boldsymbol{x}_0^k, \boldsymbol{x}_1^k, \cdots, \boldsymbol{x}_N^k, u_0^k, u_1^k, \cdots, u_N^k, s_f^k\}$.

## IV. Numerical Simulations

To validate the effectiveness of the proposed approach, performances of three trajectory planning algorithms are compared in MATLAB environment. Both r-SCP introduced here and SCP proposed in Ref. [13] are formulated and solved by the modeling toolbox YALMIP [31] and solver MOSEK [32]. These solutions are compared with the solution generated by GPOPS [33] using the default nonlinear programming solver IPOPT.

Simulation scenarios are set up mainly according to Refs. [9, 27, 34]. The initial simulation conditions are given in Table 1, and the boundary conditions are given as: $0 \leq h \leq 120$ km, $-180° \leq \theta \leq 180°$, $-180° \leq \phi \leq 180°$, $0 \leq V \leq 6,000$ m/s, $-80° \leq \gamma \leq 80°$, $-180° \leq \psi \leq 180°$, $-80° \leq \sigma \leq 80°$, $700 \leq s \leq 1,100$ km, $-10°/s \leq \dot{\sigma} \leq 10°/s$. The reference radius, gravitational acceleration, rotation rate, and atmosphere model parameters are specified as $R_0$ =3397.2 km, $g_0$ =3.7114 m/s², $\Omega$ =7.0882×10$^{-5}$ rad/s, $\rho_0$ =0.0158 kg/m³ and $h_s$ =9,354.5 m, respectively. The specifications of MSL-class vehicle is adopted, where $C_D$ =1.45, $C_L$ =0.36, $m$ =2,804 kg, $S_r$ =15.9 m², $k_Q$ =1.9027×10$^{-4}$ kg$^{0.5}$s$^{0.15}$/m$^{1.15}$, $R_n$ =6.476 m, $\dot{Q}_{max}$ =70 W/cm², $q_{max}$ =8.5 kPa, $a_{max}$ =18 $g_0$.

In the simulation, $w_{1,2,3}$, $w_v$, $w_x$, $w_s$ are set as 20, 20, 20, 5×10$^5$, 0.1, 0.1, respectively, and the convergence tolerances $\delta_{tot}$ and $v_{tot}$ are set as 10$^{-4}$ and 10$^{-5}$, which ensures that the solved trajectory satisfies the condition that the virtual control approaches zero approximately. Moreover, the initial guess trajectory for r-SCP is generated by propagating the entry dynamics Eqs. (3)~(8) with a zero initial bank angle, a constant control input $u \triangleq d\sigma/ds$ = -3, an initial guess of range 900 km, and 201 uniform discretization points. Also, the initial simulation conditions for GPOPS are given in Table 1 except $\sigma_0$, which is a free variable within its definition domain. Then the global collocation method is performed by regarding $\dot{\sigma}$ as the control input (as reported in Ref. [13]) and setting tolerance of mesh refinement as 10$^{-3}$. In addition, since only the fixed-final-time entry is investigated in Ref. [13], the flight time of SCP is prescribed as 355 seconds in cases of minimizing terminal velocity and maximizing terminal altitude. The initial guess for SCP is obtained by propagating dynamics at a constant control input $u \triangleq d\sigma/d\tau$ = -2.6 and $\sigma_0$ =0. In this way, a trajectory close to the initial guess trajectory of r-SCP is obtained for a fair comparison. This trajectory is chosen because of its good convergence for different planning problems with different performance indices while using SCP. The trust-region and stopping criterion of SCP are selected as

$$\begin{aligned}\delta &= [1000/R_0, 10\pi/180, 10\pi/180, 100/V_s, 10\pi/180, 10\pi/180, 10\pi/180] \\ \varepsilon &= [100/R_0, 0.05\pi/180, 0.05\pi/180, 10/V_s, 0.05\pi/180, 0.05\pi/180, 1\pi/180]\end{aligned} \quad (41)$$

where the two parameters $\delta$ and $\varepsilon$ are respectively the trust region radius for linearization of nonlinear dynamics and the prescribed convergence tolerance of solution. Detailed definitions and usage of these two parameters are given in Ref. [13]. Likewise, 201 uniform discretization points are also used for SCP.

It is worth noting that the SCP algorithm converges to the solution with unsatisfying accuracy in subsequent simulations because the fixed trust-region cannot guarantee the relaxation error of dynamics constraints, although this strategy is effective in the Earth entry [13]. This issue stems from sensitivity to the initial reference trajectory. In addition, the infeasible situation can be encountered during the first few iterations of the SCP in the absence of virtual control, because some final states with the initial guess may deviate largely from the expected values. The limited bank angle and its rate can also aggravate this infeasibility. To prevent the occurrence of inaccuracy and infeasibility, the virtual control technique is used in SCP to avoid the deviation of linearized dynamics, where the weight of the original performance index $w_i$ and the penalty weight $w_v$ on the virtual control in the performance index Eq. (31), are set as 1 and $10^6$, respectively.

### A. Mars entry with minimum terminal velocity

The convergence histories of altitude profiles and performance index (speed) values for minimum terminal velocity entry using r-SCP and SCP are shown in Fig. 1. The two left subplots in Fig. 1 demonstrate some intermediate profiles during the convergence process and the optimal solution marked in red. Also, the numerical solution shown in blue dash line is obtained by integrating the original nonlinear dynamics Eqs. (3)-(8) using the Runge-Kutta method with the optimal control generated by r-SCP or SCP. It can be seen that the numerical integration results are consistent with the optimal solutions, and the final flight altitude deviation is about only 50 meters for r-SCP. From the right two subplots in Fig. 1, the r-SCP converges to the optimal solution in only 7 iterations, while SCP takes 34 steps. The trajectory profiles generated by the r-SCP, SCP and GPOPS are presented in Fig. 2 and Fig. 3. As can be seen, flight states among the three methods have the same trend, which indicates that these numerical algorithms all converge to the neighborhood of the optimal solution. However, the bank angle and its rate profiles of the three approaches shown in Fig. 3 (left) are different. The bank angle and control input profiles generated by GPOPS exhibit more oscillations than those of SCP and r-SCP algorithms. In other words, the bank angle and its rate profiles of r-SCP are smoother, thus less control energy is required. The possible reason is that a smaller allowable control magnitude of $u$ than its upper bound is imposed by replacing $V\cos\gamma$ with a larger $V_{max}$ in Eq. (20) assuming a fixed maximum bank angular rate. Besides, at the beginning, bank angles of r-SCP and SCP oscillate around zero whereas the bank angle of GPOPS keeps unchanged at zero. Near the end, the bank angles of GPOPS and SCP tend to zero while the bank angle with r-SCP decreases. These differences are mainly caused by the differences in linearization, discretization and relaxation procedures. It is worth noting that the high-frequency jitter of the bank angle is avoided by choosing the bank angular rate as the system input in all the three trajectory planning methods. Additionally, as shown in Fig. 3 (right), all the path constraints are satisfied.

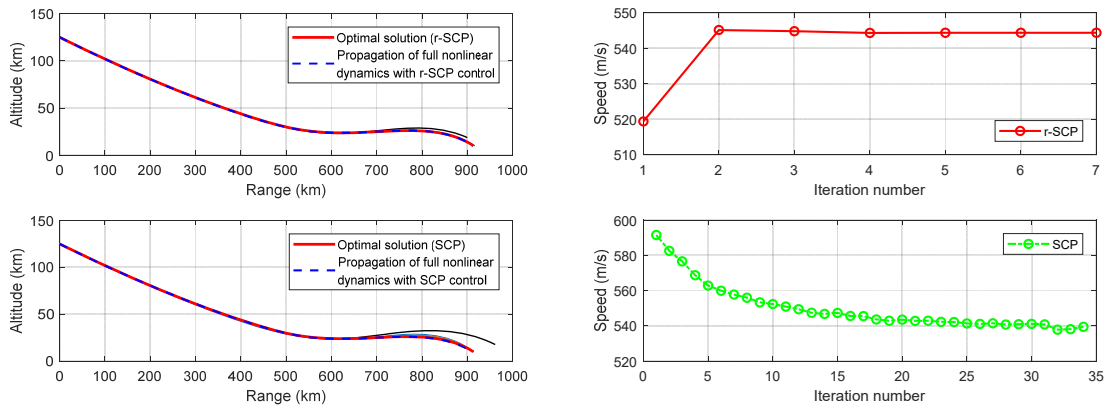

**Fig. 1 Convergence histories of altitude (left), and optimal performance index values (right) of minimum terminal velocity entry using r-SCP and SCP**

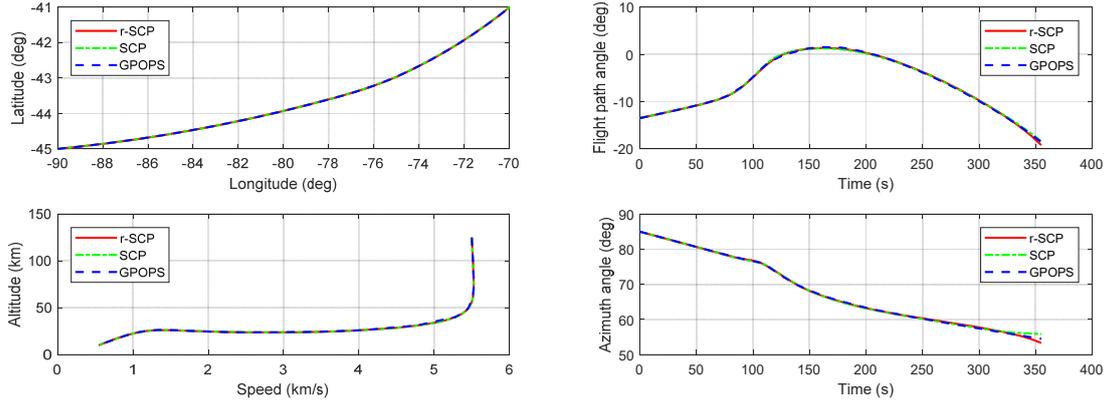

**Fig. 2 Longitude-latitude, speed-altitude profiles (left), and flight path angle and azimuth angle time histories (right) of minimum terminal velocity entry**

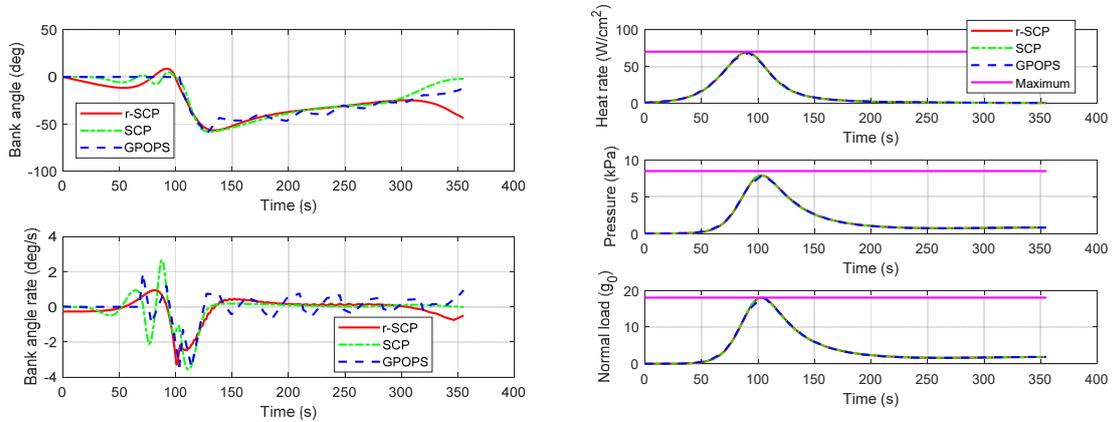

**Fig. 3 Bank angle and its rate time histories (left), and path constraints time histories (right) of minimum terminal velocity entry**

To demonstrate the performance of the algorithms quantitatively, the final states, path constraints, and CPU time consumption are summarized in Table 2. In terms of optimality, the final terminal velocity of r-SCP, SCP and GPOPS are 543.82 m/s, 539.39 m/s and 544.49 m/s, respectively. Thus, the terminal velocities of r-SCP and GPOPS are very close, and that of SCP is the minimum. Note that the optimality of r-SCP slightly degrades compared with that of SCP. As listed in Table 3, the original performance index accounts for about 94% of the new performance index for r-SCP, while the virtual control takes about 6%. For SCP, though, the original performance index accounts for nearly 100% with the virtual control only 0.68 ‰. This explains that the penalty on the virtual control for r-SCP does contribute to the change of optimality.

Table 2 Comparison of solutions for minimum terminal velocity entry

| | r-SCP | SCP | GPOPS | | r-SCP | SCP | GPOPS |
|---|---|---|---|---|---|---|---|
| $h_f$ (km) | 9.999 | 10.00 | 10 | $t_f$ (s) | 355 | 355 | 355 |
| $\theta_f$ (deg) | -70 | -70 | -70 | $\sigma_f$ (deg) | -43.49 | -1.962 | -10.71 |
| $\phi_f$ (deg) | -41 | -41 | -41 | $\left|\dot{\sigma}\right|_{max}$ (deg/s) | 2.966 | 3.580 | 3.437 |
| $V_f$ (m/s) | 543.82 | 539.39 | 544.49 | $\dot{Q}_{max}$ (W/cm$^2$) | 69.34 | 69.35 | 68.44 |
| $\gamma_f$ (deg) | -19.22 | -18.38 | -18.57 | $q_{max}$ (kPa) | 7.886 | 7.882 | 7.885 |
| $\psi_f$ (deg) | 53.36 | 55.90 | 54.53 | $n_{max}$ ($g_0$) | 17.99 | 17.99 | 18.00 |
| $s_f$ (km) | 913.18 | 913.21 | 913.26 | CPU time (s) | 4.556 | 16.58 | 11.45 |

Table 3 Components of converged new performance index

| | $w_i J_i$ | $w_v \|\bar{\boldsymbol{v}}\|_1$ | $w_x \|\bar{\delta}_x\|_2$ | $w_s \delta_s$ | $J$ |
|---|---|---|---|---|---|
| r-SCP | 3.0830 | 0.1975 | 6.1787×10$^{-5}$ | 0 | 3.2806 |
| SCP | 1.0036 | 6.8959×10$^{-5}$ | / | / | 1.0036 |

Despite optimality, r-SCP takes only 4.556 s to reach the optimal solution, whereas SCP and GPOPS take 16.58 s and 11.45 s, respectively. The trust-region radius introduced in [9] is used by r-SCP to adaptively determine the step size in each iteration and reduce the total number of optimization steps. Additionally, it should be noted that the robustness of the SCP is greatly improved by introducing the virtual control, otherwise this algorithm may not converge at the prescribed accuracy $\varepsilon$ shown in Eq. (41) even after considerable iterations. The differences of the states in the first ten iterations are shown in Table 4. Note that only the longitude, latitude and speed meet the convergence criterion. By contrast, the r-SCP and SCP with the virtual control technique converge in only a few iterations.

Table 4 Differences of the states between consecutive iterations for SCP without a virtual control

| Iteration | $|\Delta r|$ (m) | $|\Delta\theta|$ (°) | $|\Delta\phi|$ (°) | $|\Delta V|$ (m/s) | $|\Delta\gamma|$ (°) | $|\Delta\psi|$ (°) | $|\Delta\sigma|$ (°) |
|---|---|---|---|---|---|---|---|
| 1 | 6862.43 | 0.9787 | 0.1615 | 251.7393 | 4.8358 | 5.2426 | 21.8927 |
| 2 | 1501.39 | 0.0925 | 0.0594 | 91.9612 | 1.4209 | 0.7964 | 21.9099 |
| 3 | 308.74 | 0.0132 | 0.0131 | 11.7171 | 0.4698 | 0.3495 | 18.8904 |
| 4 | 471.76 | 0.0320 | 0.0441 | 14.5576 | 0.5901 | 1.14020 | 21.6865 |
| 5 | 319.97 | 0.0118 | 0.0120 | 10.6519 | 0.2693 | 0.636 | 17.7101 |
| 6 | 155.95 | 0.0078 | 0.0062 | 8.1664 | 0.1776 | 0.424 | 14.2666 |
| 7 | 124.56 | 0.0058 | 0.0038 | 4.2608 | 0.1662 | 0.2693 | 14.9714 |
| 8 | 140.13 | 0.0079 | 0.0046 | 6.3911 | 0.1146 | 0.3037 | 21.4687 |
| 9 | 115.61 | 0.0063 | 0.0032 | 3.9057 | 0.0974 | 0.275 | 16.0256 |
| 10 | 115.76 | 0.0067 | 0.0019 | 6.0361 | 0.212 | 0.6818 | 22.3167 |
| $\varepsilon$ | 100 | 0.05 | 0.05 | 10 | 0.05 | 0.05 | 1 |

## B. Mars entry with maximum terminal altitude and minimum final time

Analogously, the simulation results of maximum terminal altitude entry and minimum final time entry can be obtained. Similar to the results in previous section, the optimal trajectories generated by the three methods are nearly the same, except for slight differences in the bank angle and its rate that show smoother responses with r-SCP. Due to the space constraint, the state and control profiles and time histories are not given in figures. The terminal results of these two cases are summarized in Table 5 and Table 6, respectively. Note that only r-SCP and GPOPS are compared in Table 6 because SCP is designed for a fixed-final-time mission.

Table 5 Comparison of solutions for maximum terminal altitude entry

|  | r-SCP | SCP | GPOPS |  | r-SCP | SCP | GPOPS |
|---|---|---|---|---|---|---|---|
| $h_f$ (km) | 12.19 | 12.56 | 12.03 | $t_f$ (s) | 355 | 355 | 355 |
| $\theta_f$ (deg) | -70 | -70 | -70 | $\sigma_f$ (deg) | -45.10 | -15.53 | -11.42 |
| $\phi_f$ (deg) | -41 | -41 | -41 | $\left\|\dot{\sigma}\right\|_{max}$ (deg/s) | 6.729 | 4.379 | 10.00 |
| $V_f$ (m/s) | 626.36 | 619.54 | 605.14 | $\dot{Q}_{max}$ (W/cm$^2$) | 69.26 | 69.31 | 67.83 |
| $\gamma_f$ (deg) | -18.44 | -17.05 | -17.26 | $q_{max}$ (kPa) | 7.886 | 7.883 | 7.885 |
| $\psi_f$ (deg) | 55.30 | 58.41 | 57.46 | $n_{max}$ ($g_0$) | 17.99 | 17.99 | 18.00 |
| $s_f$ (km) | 912.44 | 912.54 | 912.73 | CPU time (s) | 5.321 | 4.886 | 5.620 |

Table 6 Comparison of solutions for minimum final time entry

|  | r-SCP | GPOPS |  | r-SCP | GPOPS |
|---|---|---|---|---|---|
| $h_f$ (km) | 9.999 | 10 | $t_f$ (s) | 332.75 | 331.08 |
| $\theta_f$ (deg) | -70 | -70 | $\sigma_f$ (deg) | -47.66 | -69.08 |
| $\phi_f$ (deg) | -41 | -41 | $\left\|\dot{\sigma}\right\|_{max}$ (deg/s) | 1.561 | 7.722 |
| $V_f$ (m/s) | 626.35 | 651.60 | $\dot{Q}_{max}$ (W/cm$^2$) | 69.34 | 67.76 |
| $\gamma_f$ (deg) | -18.65 | -20.37 | $q_{max}$ (kPa) | 7.886 | 7.845 |
| $\psi_f$ (deg) | 48.88 | 48.65 | $n_{max}$ ($g_0$) | 17.99 | 17.91 |
| $s_f$ (km) | 913.66 | 913.49 | CPU time (s) | 5.779 | 16.89 |

In Table 5, GPOPS uses the longest time (5.620 s) to find the optimal solution and generates the entry trajectory with the minimum altitude 12.03 km, while the terminal altitudes are 12.19 km with r-SCP and 12.56 km with SCP, respectively. Although the terminal altitude of r-SCP is better than GPOPS, it is not as good as the altitude using SCP. This is the price paid by the additional penalty terms $w_v \|\bar{v}\|_1 + w_x \|\bar{\delta}_x\|_2 + w_s \delta_s$ in the performance index as discussed in the previous subsection.

For the minimum time entry, r-SCP only performs 9 successive iterations to generate the converged solution, taking 11 s shorter than GPOPS. The final time achieved by GPOPS and r-SCP are 331.08 s and 332.75 s, respectively, as listed in Table 6. These results demonstrate that the optimality of r-SCP in terms of minimum final time is very close to GPOPS, while the gain in computation efficiency with r-SCP is significant.

## V. Conclusion

In this paper, the range discretization and successive convexification based approach is proposed for the Mars entry trajectory planning problem. The range is used as the independent variable to formulate the dimensionless entry dynamics rather than the energy or time, which improves the numerical accuracy of entry dynamics propagation. Another contribution of this paper is the introduction of new state and control variables in the dynamics to convexify the minimum-time performance index, as well as the inequality relaxation technique to convexify the state and control input constraints. These techniques transcribe the original non-convex Mars entry trajectory optimization problem into a convex one that can be solved in the framework of the sequential convex programming with a virtual control and an adaptive trust-region. Simulation results show that the proposed r-SCP algorithm has better performance than the conventional SCP algorithm in terms of computational efficiency and robustness to initial guess of the reference trajectory. Compared with the well-established optimization software GPOPS, r-SCP achieves comparable optimality but requires much less computation time.

## Acknowledgments

The work was supported by the National Natural Science Foundation of China (Grant No. 11672126) and Qing Lan Project. The author fully appreciates their financial supports.